\documentclass[aps,prl,twocolumn,floatfix]{revtex4}
\usepackage{amssymb,amsmath,amsfonts}
\usepackage{graphicx,graphics}
\usepackage{epsfig}
\usepackage{epstopdf}
\usepackage[FIGBOTCAP]{subfigure}
\usepackage{amsfonts}
\usepackage{bm,bbm}
\usepackage{amssymb,amsmath,amsfonts}
\usepackage{mathrsfs}
\usepackage{calc}
\usepackage{epsfig}
\usepackage{float}
\usepackage{color}
\usepackage{epstopdf}
\usepackage{bm,amssymb,amsmath}
\usepackage{graphicx,color}
\newcommand{\mez}{\hspace*{+0.50cm}}
\newcommand{\mz}{\hspace*{+0.25cm}}
\newcommand{\be}{\begin{equation}}
\newcommand{\ee}{\end{equation}}

\newcommand{\m}{\hspace*{-0.50mm}}
\newcommand{\n}{\hspace*{-0.25mm}}
\newcommand{\beqar}{\begin{eqnarray}}
\newcommand{\eeqar}{\end{eqnarray}}
\newcommand{\bcen}{\begin{center}}
\newcommand{\ecen}{\end{center}}

\begin{document}

\title{Tuning  quantum-classical correspondence of molecular systems in a cavity}

\author{Nimrod Moiseyev\footnote{{\tt nimrod@technion.ac.il}, {\tt https://nhqm.net.technion.ac.il}}}

\affiliation{Schulich Faculty of Chemistry and Faculty of Physics, Technion-Israel Institute of Technology, Haifa 32000, Israel}

\author{Milan \v{S}indelka\footnote{{\tt sindelka@ipp.cas.cz}}}

\affiliation{Institute of Plasma Physics, Academy of Sciences of the Czech Republic, Za Slovankou 1782/3, 18200 Prague 8, Czech Republic\\}

\date{\today}

\begin{abstract}
We show that the correspondence between quantum and classical mechanics can be tuned by varying the coupling strength between
an atom or a molecule and the modes of a cavity.
In the acceleration gauge (AG) representation, the cavity-matter system is described by an effective Hamiltonian, with a non-trivial coupling appearing in the potential, and with renormalized masses. Importantly, and counterintuitively, the AG coupling changes non-monotonically with the strength of the cavity-matter interaction. As a result, one obtains an effective (approximately decoupled) cavity-matter dynamics both for the case of weak and strong interactions. In the weak coupling regime, the effective mass parameters essentially coincide with their
standard interaction free counterparts. In contrast, the renormalized atomic/molecular mass increases as the cavity-matter interaction is increased.  This results in AG dynamics of matter governed by a conventionally looking atomic/molecular Hamiltonian, whose
effective Planck constant is reduced when the cavity-matter interaction is increased. This insight might lead in particular to the possibility of studying the correspondence between "quantum-chaos" (quantum stochasticity) and the classical chaos, as well as either to enhancement or to suppression of tunneling, by varying a controllable physical parameter. Physical realization of our findings is briefly discussed.
\end{abstract}

\maketitle

\section{I. Introduction and Motivation}

The correspondence between classical and quantum mechanics always attracted a lot of attention of theorists,
and in many ways it is not yet a settled issue. The classical equations of motions are non linear equations in the coordinates and momenta, whereas the quantum equation of motion (TDSE -- time dependent Schr\"{o}dinger equation) is a linear differential equation
with respect to the wavefunction.    Therefore,  chaos as it is defined  in classical mechanics  results from the non-linear nature of the classical dynamics and can not be defined and measured in quantum systems as for classical systems. As shown long ago by
Madelung \cite{MadelungREF12NMAP}, the TDSE can be transformed to two coupled  classical equations of a fluid flow without a friction and the Hamiltonian-Jacobi \cite{REF13NMAP} equation with an additional potential term which is proportional to the square of the Planck constant, $\hbar$. For any non-zero value of $\hbar$, as small as one wishes,  the linearity of the Schr\"odinger equation is camouflaged but is hidden there, and therefore the quantum dynamics comes out as obtained from the solution the time-dependent linear differential Schr\"odinger equation before the transformation.  Consequently, even in the limit of  $\hbar\to 0$, the quantum dynamics does not converge to the corresponding classical dynamics in the limit of long times.

The fact that there is no quantum analog to a single classical trajectory results mainly from the  uncertainty relation.  Indeed, as it was shown by Korsch and Berry in 1981, the "quantum map smooths out details in phase space area smaller than $\hbar$" \cite{HJKORSCHMVBERRYPhysicaD}. As it was demonstrated in Refs.~\cite{NM-Korsch,vitali,osovski}, for a driven rotor the area of the regular islands in the Poincar\'{e} surface of section and the area of the chaotic sea divided by the Planck constant gives, respectively, the  number of Floquet solutions that are localized in the regular islands, and in the closed chaotic sea. In 1983, together with Asher Peres, we have shown \cite{ASHER-NM} that quantum wavepacket does not spread at the same rate as a classical ensemble of trajectories in the same potential. Therefore, chaos seems to be well defined only in classical mechanics where one can measure the stability of a classical trajectory in phase space.\\

Stated more explicitly, the classical chaos is associated with an instability of classical trajectories, such that it is impossible to determine long term dynamics of a chaotic system even when an external perturbation on the initial point in the classical phase space is infinitesimally  small. Since in quantum mechanics there is no analog to a single classical trajectory, it seems that there is no analog to classical chaos. Therefore, rather for looking for the analog to classical chaos in quantum mechanics one should look for the fingerprints of classical chaos in quantum mechanics. See the paper of Michael Berry on "Quantum chaology, not quantum chaos", published in 1989 \cite{MBPhysScr1989}.

Another aspect of quantum  dynamics is that it introduces wave interference that is missing in classical mechanics. In 1982 and in series of papers afterwards, Shmuel Fishman and his coworkers described the strong quantum localization due to quantum interference which suppresses transport in some deterministic systems in the same way as in disordered systems (i.e., by the mechanism of Anderson Localization \cite{ShmuelFishman2010}). It is interesting to mention an experimental observation of Anderson localization of light caused by random fluctuations on a two-dimensional photonic lattice, reported in 2007 by Segev and his coworkers \cite{SchwartzTBartalGFishmanSSegevM_Nature}.

In this work, we investigate an atomic/molecular system inside an optical cavity. We show that the correspondence between quantum
and classical mechanics can be studied in such a case when the coupling between the system and the cavity modes is taken to be as
large as possible. The coupling of the quantum cavity modes with the system modes gives rise to hybridization modes known as polaritons. An extensive amount of literature (both theoretical or experimental) focuses on studying the  energy splitting
between such polaritons \cite{SPE2004,NMRbook1991,Lombarfi2007,Muller2011,Chu2009,NaturePhysics2009, Sander2012,Laraoui2010,ZollerCirac2005,small1999,YouNori2003,Reichel2004, Ladd2010,NaturePhotonics2007,LeeLee2009}.

We consider a single cavity mode with the frequency $\omega$ which is coupled with the degrees of freedom of an atomic or molecular system. The strength this coupling is denoted by $\epsilon$ in the standard momentum gauge representation.
Importantly, and counterintuitively, in the acceleration gauge representation rederived below, an increase of $\epsilon$ leads to an effective decoupling of the degrees of freedom of the system and the cavity, and the effective mass of the atomic/molecular system is increased. Moreover, we show that the resulting effective atomic/molecular Hamiltonian contains a redefined Planck constant $\hbar(\epsilon)$ which approaches zero as $\epsilon \to \infty$. Such a setup  can thus be used to study (perhaps even experimentally) the quantum classical correspondence, and in particular the fingerprints of quantum chaos.

The outline of the paper is as follows. Firstly, we rederive the acceleration gauge Hamiltonian for an atomic/molecular system
coupled to a single cavity mode of the above mentioned frequency $\omega$. Secondly, we develop a perturbation theory for the weak coupling regime where $0\le \epsilon \ll \sqrt{m\,\hbar\,\omega^3/2}$. Thirdly, we focus on the strong coupling regime, and show
that, when $\sqrt{m\,\hbar\,\omega^3/2} \ll \epsilon < \infty$, the dressed degrees of freedom of the atomic/molecular system and the cavity are
effectively decoupled, with the just mentioned dressed atomic/molecular system becoming effectively near-classical.
In the last section we conclude and emphasize on the ability to study the quantum vs.~classical chaotic dynamics in the limit of $\hbar(\epsilon) \to0$, via increasing the system-cavity coupling parameter $\epsilon$ as much as possible.

\section{II. Acceleration gauge Hamiltonian for an atomic/molecular system in a cavity}

\subsection{\sl II.A The cavity -- matter Hamiltonian in the momentum gauge (MG) representation}

We consider a model of an atomic/molecular system (represented here by a single degree of freedom) coupled to a single mode of an optical cavity. Our starting point is given by the standard momentum gauge (MG) Hamiltonian
\begin{eqnarray} \label{H-MG-def}
%  -------------------------------------------------------------------------------------------------------------------------
   & & \hat{H}_{\rm MG} \; = \\ & = & \frac{1}{2\,m} \left( \hat{p} + \frac{\epsilon}{\omega}(\hat{a}+\hat{a}^\dagger) \right)^{\m\n 2} + \; { V}\n(\hat{x}) \; + \; \hbar\omega \, \left( \hat{a}^\dagger \hat{a} \, + \, \frac{1}{2} \, \right) \; . \nonumber
%  -------------------------------------------------------------------------------------------------------------------------
\end{eqnarray}
Here $ V\n(\hat{x})$ stands for the single particle potential, $\omega$ determines the frequency of our cavity mode, and $\epsilon$ represents the "atom-cavity" coupling constant. For the sake of simplicity we have considered here a single particle Hamiltonian, yet
an extension of our discussion to many particle systems is also possible. For example, the massive particle (e.g.~an electron) can move in 3D, but for convenience we wrote explicitly only the spatial coordinate $x$ along the polarization direction of the cavity mode. % For example for a linear three body problem where two stretching modes (e.g., symmetric and anti-symmetric modes) along the $\hat x$ direction are coupled by the polarized cavity mode that is directed  along $\hat x$ as well.
Meaning of all the other symbols appearing in (\ref{H-MG-def}) should be self explanatory.
Note that the MG Hamiltonian (\ref{H-MG-def}) adopts the dipole approximation, this is legitimate whenever the wavelength
$\lambda = (2\,\pi\,c)/\omega$ is much larger than the range of ${ V}\n(x)$.

The Hamiltonian formula given in Eq.~(\ref{H-MG-def}) corresponds to a single mode approximation, where all the other cavity modes are
neglected, and contains the physical mass $m$. This approximation is limited to the case of non ultra strong coupling $\epsilon$ between the atomic/molecular system and the cavity.  In the case of an ultra strong coupling, all the cavity modes (including the highly ultraviolet ones) should be taken into consideration, and an appropriate mass renormalization procedure should be implemented,
as it appears for example in Ref.~\cite{Milonni}.
The maximum value of $\epsilon$ for which Eq.~(\ref{H-MG-def}) is a good approximation to the exact Hamiltonian is problem dependent  and we will not determine it here. Note in passing that "strong coupling is measured not by the coupling magnitude $\epsilon$ but by the observability of its consequence", see Ref.~\cite{Nitzan}.

Another useful appearance of our MG Hamiltonian, Eq.~(\ref{H-MG-def}), is obtained by employing the position and momentum operators
of the cavity oscillator, namely,
     \begin{eqnarray}
     %  -------------------------------------------------------------------------------------------------------------------------
        \hat{q} & = & i \, \sqrt{\frac{\hbar}{2\,\omega}} \, \Bigl( \hat{a} \, - \, \hat{a}^\dagger \Bigr) \mez ;\\
     %  -------------------------------------------------------------------------------------------------------------------------
        \hat{\wp} & = & \phantom{i} \, \sqrt{\frac{\hbar\omega}{2}} \, \Bigl( \hat{a} \, + \, \hat{a}^\dagger \Bigr) \mez .
     %  -------------------------------------------------------------------------------------------------------------------------
     \end{eqnarray}
     Observables $\hat{q}$ and $\hat{\wp}$ satisfy the standard commutation property $[\hat{q},\hat{\wp}]=i\hbar\,\hat{1}$.
     Instead of (\ref{H-MG-def}) one may write now simply
     \begin{eqnarray} \label{H-MG-def-q-p}
     %  -------------------------------------------------------------------------------------------------------------------------
        & & \hat{H}_{\rm MG} \; = \\ & = & \frac{1}{2\,m} \Bigl( \hat{p} + \varsigma\,\hat{\wp} \Bigr)^{\m\m 2} + \;
        {V}\n(\hat{x}) \; + \; \frac{1}{2} \, \hat{\wp}^2 \; + \; \frac{1}{2} \, \omega^2 \, \hat{q}^2 \mz ; \nonumber
     %  -------------------------------------------------------------------------------------------------------------------------
     \end{eqnarray}
     where by definition
     \be \label{varsigma-def}
        \varsigma \; = \; \frac{\epsilon}{\omega} \, \sqrt{\frac{2}{\hbar\omega}} \mez .
     \ee
     Note that, as $\epsilon$ increases, the MG coupling between the cavity and the system is increased as well.
     As we will show below, the situation in the acceleration gauge representation is, rather surprisingly,
     qualitatively different and much more physically interesting.

\subsection{\sl II.B Transforming the cavity -- matter MG Hamiltonian into the acceleration gauge (AG) representation}

An unitary operator
     \be \label{hat-U-MA}
        \hat{U}_{\rm MA} \; = \;
        e^{+\frac{\epsilon\,\omega\,(\hat{a}-\hat{a}^\dagger)\,\hat{p}}{m\,\hbar\,\omega^3 \, + \, 2\,\epsilon^2}} \; = \;
        e^{-i\frac{\epsilon\,\omega\,\sqrt{2\omega/\hbar}\,\hat{q}\,\hat{p}}{m\,\hbar\,\omega^3 \, + \, 2\,\epsilon^2}}
     \ee
     generates an equivalent AG Hamiltonian
     \begin{eqnarray} \label{H-AG-def}
     %  --------------------------------------------------------------------------------------------------------------------------
        \hspace*{-0.35cm} & & \hat{H}_{\rm AG} \; = \\
        \hspace*{-0.35cm} & = & \hat{U}_{\rm MA}^\dagger \, \hat{H}_{\rm MG} \, \hat{U}_{\rm MA} \; = \nonumber\\
     %  -------------------------------------------------------------------------------------------------------------------------
        \hspace*{-0.35cm} & = & \frac{\hat{p}^2}{2\,M(\epsilon)} \; + \; V\m\Bigl(\hat{x}+\zeta(\epsilon)\,\hat{q}\Bigr) \; + \;
        \frac{\hat{\wp}^2}{2\,\mu(\epsilon)} \; + \; \frac{1}{2} \, \mu(\epsilon) \, \Omega^2\n(\epsilon) \, \hat{q}^2 \; . \nonumber
     %  --------------------------------------------------------------------------------------------------------------------------
     \end{eqnarray}
     Here by definition
     \be \label{M-lambda}
        M(\epsilon) \; = \; \, \left(\frac{m\,\hbar\,\omega^3\,+\,2\,\epsilon^2}{\hbar\,\omega^3}\right) \mez ;
     \ee
    \be \label{zeta-lambda}
        \zeta(\epsilon) \; = \; \frac{\epsilon\,\omega\,\sqrt{2\,\hbar\omega}}{m\,\hbar\,\omega^3 \, + \, 2\,\epsilon^2} \mez ;
     \ee
     \be \label{mu-lambda}
        \mu(\epsilon) \; = \; \frac{m\,\hbar\,\omega^3}{m\,\hbar\,\omega^3 \, + \, 2\,\epsilon^2} \; = \; \frac{m}{M(\epsilon)} \mez ;
     \ee
     \be \label{Omega-lambda}
        \Omega(\epsilon) \; = \; \frac{\omega}{\sqrt{\mu(\epsilon)}} \mez .
     \ee
     Validity of Eq.~(\ref{H-AG-def}) can be straightforwardly verified, given the fact that the unitary operator $\hat{U}_{\rm MA}$
     of Eq.~(\ref{hat-U-MA}) generates a spatial translation along $\hat{x}$ and a momentum translation along $\hat{\wp}$. That is,
     \begin{eqnarray}
     %  --------------------------------------------------------------------------------------------------------------------------
        \hat{U}_{\rm MA}^\dagger \, \hat{x} \, \hat{U}_{\rm MA} & = & \hat{x} \; + \; \zeta(\epsilon) \, \hat{q} \mez ;\\
     %  --------------------------------------------------------------------------------------------------------------------------
        \hat{U}_{\rm MA}^\dagger \, \hat{\wp} \, \hat{U}_{\rm MA} & = & \hat{\wp} \; - \; \zeta(\epsilon) \, \hat{p} \mez .
     %  --------------------------------------------------------------------------------------------------------------------------
     \end{eqnarray}
     The Hamiltonian formula (\ref{H-AG-def}) agrees with an expression derived in Ref.~\cite{Bandrauk} by the so called
     Bloch-Nordsieck transform for the case of many electron atoms/molecules interacting with a multimode quantum electromagnetic field.

     In the AG representation, the coupling between our atomic/molecular system and the cavity enters solely into the potential
     term $V\m\Bigl(\hat{x}+\zeta(\epsilon)\,\hat{q}\Bigr)$. This coupling is {\it not}
     a function of $\epsilon$ as in the MG representation, but a function of another parameter $\zeta(\epsilon)$ which depends non-monotonically upon $\epsilon$, see Eq.~(\ref{zeta-lambda}). For small values of $\epsilon$, the parameter $\zeta(\epsilon)$ increases as $\epsilon$ is increased. As $\epsilon$ grows further, $\zeta(\epsilon)$ reaches its maximum value (i.e., the strongest AG coupling between the atomic/molecular system and the cavity), corresponding to
     \be
        \epsilon_{\max} \; = \; \sqrt{\frac{m\,\hbar\,\omega^3}{2}} \mez , \mez
        \zeta(\epsilon_{\max}) \; = \; \frac{1}{2\,\sqrt{m}} \mez .
     \ee
     When increasing $\epsilon$ further beyond $\epsilon_{\max}$, the AG coupling $\zeta(\epsilon)$ gradually falls off to zero.
     We will return to this interesting observation later on in section III.B.

     Recall again that the system -- cavity coupling is in the AG representation contained solely in the displaced atomic potential ${ V}\m\Bigl(\hat{x}+\zeta\,\hat{q}\Bigr)$. Therefore, the AG coupling is restricted only to a finite spatial
     region determined by the range of ${ V}$\m. In other words, \textit{the atom-field coupling vanishes outside the cavity, this is
     the basic advantage of AG representation, showing that the AG lends itself very well to the description of ionization and scattering phenomena.} Importantly, in the AG, the original particle mass $m$ is renormalized to $M(\epsilon)$ of Eq.~(\ref{M-lambda}). Also the mass of our field oscillator is renormalized here to $\mu(\epsilon)$ of Eq.~(\ref{mu-lambda}), and its frequency $\omega$ to $\Omega(\epsilon)$ of Eq.~(\ref{Omega-lambda}). In other words,
     \textit{\textit{both the atomic/molecular system and the cavity oscillator become dressed by the interaction}}.

\section{III. The weak and the strong coupling regimes}

      Perturbative treatment of the problem defined by the Hamiltonian (\ref{H-AG-def}) implies Taylor series expansion of
      $\hat{H}_{\rm AG}$ in a small coupling parameter $\zeta(\epsilon)$ of Eq.~(\ref{zeta-lambda}). Clearly, $\zeta(\epsilon)$ is an analytic function
      of the MG coupling $\epsilon$ in two distinct situations. Either when  $2\,\epsilon^2 \ll m\,\hbar\,\omega^3$, then $\zeta(\epsilon)$ is linearly proportional to $\epsilon$. Or as long as $2\,\epsilon^2 \gg m\,\hbar\,\omega^3$, then $\zeta(\epsilon)$ is linearly proportional to the \textit{inverse} of  $\epsilon$. Therefore, a very strong MG coupling $\epsilon$ implies in this situation a small value of the AG coupling $\zeta(\epsilon)$, as already pointed out above. Let us look now at the two just mentioned distinct regimes in some more detail.\\

\subsection{\sl III.A The weak coupling regime: $0\le \epsilon \ll \sqrt{m\,\hbar\,\omega^3/2}$}

In the weak coupling regime of small $\epsilon$, the MG formulas (\ref{H-MG-def}) and (\ref{H-MG-def-q-p}) reduce to
     \begin{eqnarray} \label{MG-WEAK}
     %  -------------------------------------------------------------------------------------------------------------------------
        \hspace*{-0.50cm} & & \hat{H}_{\rm MG} \; = \\ \hspace*{-2.00cm} & = & \frac{\hat{p}^2}{2\,m} \; + \; V\m(\hat{x}) \; + \;
        \hbar\omega \, \left( \hat{a}^\dagger \hat{a} \, + \, \frac{1}{2} \, \right) \nonumber\\
        \hspace*{-0.50cm} & + &
        \frac{\xi}{m\,\omega} \; \hat{p} \, \Bigl( \hat{a} + \hat{a}^\dagger \Bigr) \; + \; {\cal O}(\xi^2) \; = \nonumber\\
     %  -------------------------------------------------------------------------------------------------------------------------
        \hspace*{-0.50cm} & = & \frac{\hat{p}^2}{2\,m} \; + \; V\m(\hat{x}) \; + \; \frac{1}{2} \, \hat{\wp}^2 \; + \;
        \frac{1}{2} \, \omega^2 \, \hat{q}^2 \; + \; \frac{\varsigma}{m} \, \hat{p} \, \hat{\wp} \; + \; {\cal O}(\varsigma^2) \; . \nonumber
     %  -------------------------------------------------------------------------------------------------------------------------
     \end{eqnarray}
     Hence the leading order matter -- cavity coupling term equals to $\frac{\varsigma}{m} \, \hat{p} \, \hat{\wp}$ where
     $\varsigma$ is given by Eq.~(\ref{varsigma-def}).

     The AG formula (\ref{H-AG-def}) reduces to
     \begin{eqnarray}
     %  -------------------------------------------------------------------------------------------------------------------------
     \label{AG-WEAK}
        \hspace*{-0.50cm} & & \hat{H}_{\rm AG} \; = \\ \hspace*{-0.50cm} & = & \frac{\hat{p}^2}{2\,m} \; + \; V\m(\hat{x}) \; + \;
        \frac{1}{2} \, \hat{\wp}^2 \; + \; \frac{1}{2} \, \omega^2 \, \hat{q}^2 \; + \; V'\n(\hat{x}) \, \zeta \, \hat{q}
        \; + \; {\cal O}(\zeta^2) \; . \nonumber
     %  -------------------------------------------------------------------------------------------------------------------------
     \end{eqnarray}
     Hence the leading order matter -- cavity coupling term equals to $V'\n(\hat{x}) \, \zeta \, \hat{q}$ where
     $\zeta(\epsilon)$ is given by Eq.~(\ref{zeta-lambda}).

     Equations (\ref{MG-WEAK}) and (\ref{AG-WEAK}) show that, in the weak coupling regime of small $\epsilon$, one might expect that the deviation of the quantum dynamics from the classical one will increase as the system is placed in the cavity (as for example by trapping the atomic/molecular system in between two mirrors). This applies in particular for isomerisation  reactions, or when a molecular system is initially prepared in a predissociation quasi-bound state. In both cases tunneling is the key role mechanism, and the tunneling rates are expected to increase with coupling the molecule to a cavity mode as displayed in (\ref{MG-WEAK}) and (\ref{AG-WEAK}). Such an anticipated effect of the cavity on the classical-quantum correspondence turns out to be very different from the situation encountered in the strong coupling regime ($\epsilon$ large) which is discussed below in III.B.

\subsection{\sl III.B The strong coupling regime: $\sqrt{m\,\hbar\,\omega^3/2} \ll \epsilon$}

Our discussion of the regime of large $\epsilon$ becomes most transparent after introducing an additional rescaling transformation
     \begin{eqnarray}
     %  --------------------------------------------------------------------------------------------------------------------------
        \label{hat-Q-def} \hat{q} & = & \left(\sqrt{\mu(\epsilon)}\;\omega\right)^{\m\m-\frac{1}{2}} \hat{{\cal Q}} \mez ;\\
     %  --------------------------------------------------------------------------------------------------------------------------
        \label{hat-P-def} \hat{\wp} & = & \left(\sqrt{\mu(\epsilon)}\;\omega\right)^{\m\m+\frac{1}{2}} \hat{\cal P} \mez .
     %  --------------------------------------------------------------------------------------------------------------------------
     \end{eqnarray}
The redefined observables $\hat{{\cal Q}}$ and $\hat{\cal P}$ possess again the standard commutation property
$\Bigl[\hat{{\cal Q}},\hat{\cal P}\Bigr]=i\hbar\,\hat{1}$. Combination of (\ref{hat-Q-def})-(\ref{hat-P-def}) and (\ref{H-AG-def})
yields accordingly
     \begin{eqnarray} \label{H-AG-def-2}
     %  --------------------------------------------------------------------------------------------------------------------------
        \hspace*{-0.35cm} & & \hat{H}_{\rm AG} \; = \\
     %  -------------------------------------------------------------------------------------------------------------------------
        \hspace*{-0.35cm} & = & \frac{\hat{p}^2}{2\,M(\epsilon)} \; + \; V\m\Bigl(\hat{x}+\xi(\epsilon)\,\hat{{\cal Q}}\Bigr) \; + \;
        \Omega(\epsilon) \left( \frac{\hat{\cal P}^2}{2} \; + \; \frac{\hat{\cal Q}^2}{2} \right) \; ; \nonumber
     %  --------------------------------------------------------------------------------------------------------------------------
     \end{eqnarray}
where by definition
\be \label{xi-def}
   \xi(\epsilon) \; = \; \left(\sqrt{\mu(\epsilon)}\;\omega\right)^{\m\m-\frac{1}{2}} \zeta(\epsilon) \mez .
\ee

Let us begin analyzing now the Hamiltonian formula (\ref{H-AG-def-2}) in the regime of large $\epsilon$. Equations
(\ref{zeta-lambda}) and (\ref{mu-lambda}) imply that $\zeta(\epsilon)$ and $\mu(\epsilon)$ are for large $\epsilon$ of orders ${\cal O}(\epsilon^{-1})$ and ${\cal O}(\epsilon^{-2})$, respectively. If so, then $\xi(\epsilon)$ of Eq.~(\ref{xi-def}) is at large $\epsilon$ of ${\cal O}(\epsilon^{-1/2})$, whereas $\Omega(\epsilon)$ of Eq.~(\ref{Omega-lambda}) is of ${\cal O}(\epsilon^{+1})$. The just presented order-of-magnitude estimates lead to striking consequences: Since, at large values of $\epsilon$, the parameter $\xi(\epsilon)$ vanishes while $\Omega(\epsilon)$ diverges to infinity, {\it the dressed (AG) degrees of freedom of our atomic/molecular system become decoupled from the dressed cavity oscillator.} Simultaneously, {\it the dressed cavity oscillator remains effectively in its ground state}, while merely lifting
the eigenvalues of $\hat{H}_{\rm AG}$ by its zero point energy contribution $\hbar\Omega(\epsilon)/2$. In passing we note that
the dressed cavity coordinate ${\cal Q}$ remains spatially confined, because our dressed cavity oscillator stays almost
unexcited as just emphasized above.

Furthermore, in the regime of large $\epsilon$, the dressed atomic/molecular mass $M(\epsilon)$ of Eq.~(\ref{M-lambda}) diverges to infinity as ${\cal O}(\epsilon^{+2})$. Equivalently one may say that the effective Planck constant $\hbar_{\rm eff}(\epsilon)=\hbar\,\sqrt{m/M(\epsilon)}$ falls off to zero as ${\cal O}(\epsilon^{-1})$.
This implies however that {\it the dressed atomic/molecular Hamiltonian
\begin{eqnarray} \label{H-AG-S}
   \hspace*{-1.00cm} \hat{H}_{\rm AG}^{\rm S}(\epsilon\to \infty) & = & \frac{\hat{p}^2}{2\,M(\epsilon)} \; + V\m(\hat{x}) \; = \nonumber\\
   \hspace*{-1.00cm} & = & -\,\frac{\hbar_{\rm eff}^2\n(\epsilon)}{2\,m} \, \partial_{xx} \; + \; V\m(x)
\end{eqnarray}
describes for $\epsilon \to \infty$ a nearly classical (semiclassical) system.} Here we obtain a spectacular result!
We recall in this context that $\epsilon$ is a controllable physical parameter.

We recall however that $\epsilon$ cannot be taken extremely large due to the issue of mass renormalization pointed out above ({\sl cf.}~Ref.~\cite{Milonni}). On the other hand, we argue that the Hamiltonian of Eq.~(\ref{H-AG-S}) might perhaps motivate even
experimental studies of quantum-classical correspondence, performed in the regime when $\epsilon$ is taken to be large yet our Hamiltonians (\ref{H-MG-def}) and (\ref{H-AG-S}) still provide a physically adequate description.

% Although , it is not immediately obvious here that $\zeta(\epsilon)\hat q\to 0$ since $-\infty\le q\le+\infty$. However, in semiclassical approximation of quantum systems the eigenstates often are confined to finite (allowed) regions in phase space where the energy is bounded the approximation leading to Eq.\ref{H-AG-def-3} is valid for a case of  $q$ values, and one can expect good quantum-classical correspondence for states of the atoms/molecules that are supported in these $q$ values.

%{\color{magenta} {\color{red} NOT NEEDED} Let us be more specific when  the %cavity is described as harmonic oscillator with the quantized energy %$E_{cav}(n_q)=\hbar \omega \sqrt{\frac{M(\epsilon)}{m(n_q+1/2)}}$. Therefore, %$$q_{rms}=\sqrt{\frac{\hbar}{(n_q+1/2)}}\frac{1}{\omega}$$ (since half of %$E_{cav}$ is the mean potential energy of harmonic oscillator) where $rms$ %stands for root-mean-square. The small parameter in the perturbation expansion %is $\zeta(\epsilon)q_{rms}$ which is vanished as $\zeta(\epsilon) \sim %1/\epsilon \to 0$ when $\epsilon\to\infty$ as long as the energy of the cavity %is confined to $0\le n_q\le N_q$ where $E_{cav}(N_q)=constant$.

%Consequently when $\epsilon\to \infty$ the potential %$V(x+\zeta(\epsilon)q_{rms})\to V(x)$ and therefore

%     the Hamiltonian is  almost a separable operator which is weakly dependent %on $\omega$.}

We have thus shown that
\textit{the correspondence between quantum to classical mechanics can be studied for atomic/molecular systems in a cavity when the coupling between the system and the cavity modes is taken to be as large as possible due to the current technology limitations and due to the validity of our system -- cavity model Hamiltonian of Eq.~(\ref{H-MG-def}) as discussed here. This kind of studies is out of the scope of the present work.} However, as we will point out in the next concluding section, our finding may even at its current stage
open a dialogue between theoreticians and experimentalists that might result in fruitful collaborations.

\section{IV. Concluding remarks}

In the present article, we have investigated an atomic/molecular system coupled to a single mode of an optical cavity.
Mixing of the atomic/molecular degrees of freedom with the cavity mode gives rise to the so called polaritons.
Within the framework of the AG representation, the Hamiltonian of the polaritons consists of a kinetic energy
operator that is separable in the dressed atomic/molecular and the dressed cavity coordinates, and a non-separable potential
depending upon these dressed atomic/molecular
and cavity coordinates. The relevant AG coupling, $\zeta(\epsilon)$ (or equivalently $\xi(\epsilon)$), is a function of the standard MG coupling parameter
$\epsilon$. The masses that appear in the separable kinetic energy operators are also functions of $\epsilon$.  For a sufficiently large value of $\epsilon$, the dressed atomic/molecular system becomes almost separable from the dressed cavity mode, and
the effective mass of the atomic/molecular system is increased. Hence the associated effective Planck constant is reduced.
We suggest that this insight enables possible studies of dynamics and spectra of quantum systems as one approaches the classical regime
for large values of $\epsilon$.

One possible example of application may concern studying the fingerprints of classical chaos in the H\'{e}non-Heiles Hamiltonian (which consists of symmetrical and asymmetrical stretching modes that  are non-linearly coupled  one to another). Note that the quantum vs.~classical dynamics of H\'{e}non-Heiles Hamiltonian was studied  extensively through the last four decades (see for example  Ref.~\cite{ASHER-NM}). \textit{The physical realization} of the H\'{e}non-Heiles model might correspond for example to a study of the dynamics and vibrational spectroscopy of $CO_2$ in a cavity.

 Before continuing with other possible applications, we wish to emphasize that the coupling parameter $\epsilon^2$ is linearly proportional to the inverse of the volume of the cavity, $V$. This volume can be even smaller than $1\;{\rm nm}^3$ as explained in Science paper from 2016 \cite{science2016}. Therefore,  $1/V=10^{27}\,{\rm cm}^{-3}$. So our proposed experiments seem to be doable when plasmonic cavities (“picocavities”) are used.

%  When mirror cavities are used then the number of molecules in the cavity enhances the strength of the coupling. In two-level model per molecule it is shown to be increased by the square root of the number of molecules that can be on the order of Avogadro number.  See the proof for two-level system approximation in the review of Nitzan and coauthors\cite{Nitzan}. Here below  our proof for the M-level molecular Hamiltonian , ${\bf H}_M$ (with the $MxM$ dimension) and N is the number of the molecules in the cavity.
%
%  The dressed Hamiltonian for N molecules in the cavity is given by an $[M(N+1)]x[M(N+1)]$ matrix, ${\bf H}_{dress}$. The non-zero matrix elements are  the diagonal block matrices,  $H_{dress}(1,1)={\bf H}_M+\hbar\omega {\bf I}$ due the absorption of one photon and  $H_{dress}(i=1,..,N,j=i)={\bf H}_M$, and the off-diagonal block matrices (first row and column of the super matrix  ${\bf H}_{dress}$) are given by $\epsilon {\bf H}_{coupl}/\sqrt{N}$ (normalization by number of molecules in the cavity). Simple algebra shows that the spectrum of the dressed N-level molecular system consists of N degenerated spectrum of the free-cavity molecular system, ${\bf H}_M$, which are the non-zero values of the diagonal matrix ${\bf E}_M$. That is, $ {\bf H}_M {\bf \Psi}_M={\bf \Psi}_M{\bf E}_M$ and therefore, $${\bf E}_M={\bf \Psi}_M^T {\bf H}_M {\bf \Psi}_M$$; and of the eigenvalues of the matrix $$
%  {\bf E}^{polaritons}={\bf E}_M+\sqrt{N}\epsilon{\bf \Psi}_M^T {\bf H}_{coupl} {\bf \Psi}_M$$.

\textit{Another physical realization} might concern the suppression of tunneling in the semiclassical limit.  Laser spectroscopy studies of diatomic molecules in a cavity (for different values of $\epsilon$) are expected to show the reduction of the dissociation lifetime (i.e., suppression of the tunneling rate). Another example is to study an effect of the cavity on shape type resonances of molecular anions which become more stable (acquire a longer lifetime) as effective $\hbar(\epsilon)$ is reduced.
One may think here e.g.~of molecular anions such as molecular nitrogen anion \cite{nitrogenANION} in its ground state, or uracil anion in its ground and excited states which we studied recently \cite{uracilANION}, or exotic atomic anions such as positronium negative anion that were observed in bombardment of slow positrons onto a Na-coated W surface \cite{positroniumANION}.
Note however that, based on our above derivation for the weak coupling regime, as $\epsilon$ is increased the coupling between the bound states and resonances with the continuum of the molecules will be increased. Therefore, we expect that in this weak coupling regime the resonance decay rate (inverse lifetime) will be increased rather than decreased, contrary to the strong coupling regime.\\

\acknowledgments{ It is a great pleasure to dedicate this article to Michael Berry on the occassion of his 80-th birthday.
%It is a great pleasure to contribute an article to this special issue which is dedicated to the 80-th birthday of Michael Berry, a dear friend and colleague, who  keeps his enthusiasm for science, his creativity and originality. His personality and enthusiasm for science  have been a role model to young people who chose  a carrier  in science.
Prof. Saar Rahav from the Technion is acknowledged for most helpful enlightening comments. The Israel Science Foundation (Grant No.~1661/19) is acknowledged for a partial support.
M.~v{S}.~acknowledges financial support of the Grant Agency of the Czech Republic (grant No.~20-21179S).}

\end{document}